%% file: main.tex
\documentclass[sigconf, authorversion, screen]{acmart}

\usepackage{graphicx}
\usepackage[T1]{fontenc}
\usepackage[utf8]{inputenc}
\usepackage{xcolor}
\usepackage{tcolorbox}

\usepackage{booktabs}
\usepackage{float}
\usepackage{subcaption}
\usepackage{tikz}
\usepackage{xparse}

\usepackage{threeparttable}
\usepackage{multirow}
\usetikzlibrary{calc}
\tcbuselibrary{skins}

\input{diagrams/kaviat_diagrams/kaviat_1}

\title{Spec2COBOLRot: An Agentic-AI Degradation Loop for Realistic COBOL Corpus Generation}

\author{Jean-Baptiste Espinasse}
\orcid{0009-0006-2569-3992}
\affiliation{%
  \institution{Sopra Steria}
  \city{Nantes}
  \country{France}
}
\affiliation{%
  \institution{Inria}
  \city{Rennes}
  \country{France}
}
\email{jb.espinasse@soprasteria.com}

\author{Djamel Eddine Khelladi}
\orcid{0000-0002-2218-650X}
\affiliation{%
  \institution{CNRS, Univ. of Rennes, IRISA, Inria}
  \city{Rennes}
  \country{France}
}
\email{djamel-eddine.khelladi@irisa.fr}

\author{Mathieu Acher}
\orcid{0000-0003-1483-3858}
\affiliation{%
  \institution{INSA, Univ. of Rennes, IRISA, Inria}
  \city{Rennes}
  \country{France}
}
\email{mathieu.acher@irisa.fr}

\date{August 2026}

\setcopyright{cc}
\setcctype{by-nc-nd}
\acmDOI{10.1145/3843775.3844543}
\acmYear{2026}
\copyrightyear{2026}
\acmISBN{979-8-4007-2987-4/2026/10}
\acmConference[AISM '26]{Proceedings of the 2nd International Workshop on AI for Software Modernization}{October 12--16, 2026}{Munich, Germany}
\acmBooktitle{Proceedings of the 2nd International Workshop on AI for Software Modernization (AISM '26), October 12--16, 2026, Munich, Germany}
\acmSubmissionID{asews26aismmain-p2-p}
\received{2026-08-11}
\received[accepted]{2026-08-23}

\begin{abstract}
    COBOL remains widely deployed, % in mission-critical systems, 
    yet
    representative corpora reflecting real production code are rarely available, % in open access, 
    limiting rigorous benchmarking of modernization approaches.
    We propose a systematic agentic AI pipeline for generating realistic COBOL
    programs, combining specification-driven generation with iterative degradation
    guided by patterns and complexity targets extracted from real production code. 
    Here, realism is understood as structural fidelity to production code  as captured by our metrics.
    We evaluate whether degradation reaches target complexity levels while
    preserving business behavior, and examine the limits of the approach,
    across three programs from distinct business domains. Results show the pipeline
    reliably produces syntactically valid programs and moves them toward realistic
    structural complexity. However, preserving business behavior %exactly 
    is not always
    achieved by construction, and %, more fundamentally, 
    targeting structural
    metrics independently of business logic risks producing programs whose
    complexity does not reflect a plausible maintenance history. We discuss these
    limitations and outline a more realistic alternative as a direction for future
    work, generating legacy programs from scratch along a simulated development
    history. %, as a promising direction for futur work
\end{abstract}

\keywords{Automated Software Engineering, COBOL Generation, Large Languages Models}

\ccsdesc[500]{Software and its engineering~Software maintenance tools}

\begin{document}
    \maketitle

    \input{sections/introduction}

    \input{sections/related_work}

    \input{sections/corpus_analysis}

    \input{sections/approach}

    \input{sections/experimental_setup}

    \input{sections/evaluation}

    \input{sections/threats}

    \input{sections/discussion}

    \input{sections/conclusion}

    \bibliographystyle{ACM-Reference-Format}
    \bibliography{references}
\end{document}

%% file: diagrams/kaviat_diagrams/kaviat_1.tex
\NewDocumentCommand{\KiviatLegend}{}{%
\begin{tikzpicture}[baseline]
    \fill[green!45, fill opacity=0.5] (0,0) rectangle (0.6,0.3);
    \draw[very thick, dash pattern=on 6pt off 4pt, draw=green!50!black]
        (0,0) -- (0.6,0);
    \node[font=\footnotesize, anchor=west] at (0.7,0) {Target band (min $\rightarrow$ max)};

    \draw[very thick, draw=red!70!black] (5.6,0) -- (6.2,0);
    \node[font=\footnotesize, anchor=west] at (6.3,0) {Pre-degradation};

    \draw[very thick, draw=blue!80!black] (8.8,0) -- (9.4,0);
    \node[font=\footnotesize, anchor=west] at (9.5,0) {Post-degradation};
\end{tikzpicture}%
}

\NewDocumentCommand{\KiviatDiagram}{mmm}{%
\begin{tikzpicture}[every node/.style={align=center}]
    \def\Rmax{4}
    \def\Vmax{5}
    \pgfmathsetmacro{\Step}{360/7}

    \foreach [count=\i] \lab in {SLOC,NPAR,CC,APL,DIVR,NGO,VUR}{
        \pgfmathsetmacro{\ang}{90-(\i-1)*\Step}
        \node[font=\Large] at (\ang:{\Rmax+1.1}) {\lab};
    }

    \foreach \r in {1,...,5}{
        \def\GridPath{}
        \foreach \i in {0,...,6}{
            \pgfmathsetmacro{\ang}{90-\i*\Step}
            \xdef\GridPath{\GridPath (\ang:{\r*\Rmax/5}) --}
        }
        \draw[gray!35, thin] \GridPath cycle;
    }
    \foreach \r/\lab in {1/1,2/2,3/3,4/4,5/5}{
        \node[gray!60, font=\tiny, anchor=west] at ({90-3}:{\r*\Rmax/5}) {\lab};
    }
    \foreach \i in {0,...,6}{
        \pgfmathsetmacro{\ang}{90-\i*\Step}
        \draw[gray!55] (0,0) -- (\ang:\Rmax);
    }

    \def\PathBandOuter{}
    \foreach \i in {0,...,6}{
        \pgfmathsetmacro{\ang}{90-\i*\Step}
        \xdef\PathBandOuter{\PathBandOuter (\ang:\Rmax) --}
    }
    \def\PathBandInner{}
    \foreach [count=\i] \v in {#1}{
        \pgfmathsetmacro{\ang}{90-(\i-1)*\Step}
        \xdef\PathBandInner{\PathBandInner (\ang:{\v/\Vmax*\Rmax}) --}
    }
    \path[draw=none, fill=green!45, fill opacity=0.20, even odd rule]
        \PathBandOuter cycle
        \PathBandInner cycle;
    \draw[very thick, dash pattern=on 6pt off 4pt, draw=green!50!black]
        \PathBandInner cycle;

    \def\PathInit{}
    \foreach [count=\i] \v in {#2}{
        \pgfmathsetmacro{\ang}{90-(\i-1)*\Step}
        \xdef\PathInit{\PathInit (\ang:{\v/\Vmax*\Rmax}) --}
        \filldraw[red!70!black] (\ang:{\v/\Vmax*\Rmax}) circle (1.8pt);
    }
    \draw[very thick, draw=red!70!black, fill=red!60, fill opacity=0.12]
        \PathInit cycle;

    \def\PathFin{}
    \foreach [count=\i] \v in {#3}{
        \pgfmathsetmacro{\ang}{90-(\i-1)*\Step}
        \xdef\PathFin{\PathFin (\ang:{\v/\Vmax*\Rmax}) --}
        \filldraw[blue!80!black] (\ang:{\v/\Vmax*\Rmax}) circle (1.8pt);
    }
    \draw[very thick, draw=blue!80!black, fill=blue!40, fill opacity=0.12]
        \PathFin cycle;

\end{tikzpicture}%
}

%% file: sections/introduction.tex
\section{Introduction}
\label{sec:intro}

% Presence and adoption of COBOL
Legacy systems are running critical business activity across banking, insurance,
public administration, etc, where reliability is important. %non-negotiable
%Among the technologies that have sustained these mainframe environments, 
COBOL remains one
of the most widely deployed programming languages in production systems, with
billions of lines still actively executed today~\cite{caballar2025ibmcobolmodernization}. Originally designed for business applications, COBOL was adopted for its readability relative to today languages, making it accessible to a broad range of practitioners.

% Complexity, drift over year
However, the characteristics that once made COBOL well-suited for its context
have become sources of difficulty. Compared to modern languages, COBOL systems
are increasingly hard to maintain, the generation of developers who originally built
these systems is retiring, leaving entry-level developers to face significant
challenges in code comprehension and defect location~\cite{Ciborowska_2021, lee2022,
dau2026}. Beyond the workforce transition, the software engineering paradigms
and conventions applied during the development of these systems differ
substantially from modern practices. Code that was considered well-structured at
the time has accumulated decades of patches, feature additions, and emergency fixes,
often without corresponding updates to documentation. This progressive drift between
code and its specifications has produced large and opaque codebases. % that are critical to operations yet poorly understood, even by those who maintain them.
Structured programming standards such as Jackson Structured Programming~\cite{jsp}
were introduced to address some of these issues, but their adoption was uneven,
and the accumulation of technical debt has continued. Furthermore, these legacy
technologies are largely absent from contemporary academic curricula, further widening
the skills gap. 
%
% New usage of GenAI for modernization use case.
Recent work has explored generative AI and agentic AI for COBOL modernization tasks such
as retro-documentation, business rules extraction and code transformation \cite{dau2026,lei2025enhancingcobolcodeexplanations,chiranjeevi2025,ghandi2024},
showing promise over %traditional 
rule-based approaches that operate at the syntactic
level and therefore fail to capture business intent\cite{chiranjeevi2025}. However, 
%these approaches 
they come with significant limitations. LLM outputs are non-deterministic,
opaque, difficult to control, and the resulting translations cannot be trusted without
extensive validation \cite{hans2025automatedtestingcoboljava,kumar2025automatedvalidationcoboljava,froimovich2025qualityevaluationcoboljava}. 
%
% Evaluation has never been more important
Given that these legacy systems have sustained mission-critical operations for
decades, any modernization effort introduces substantial risk. This concern is
widely shared among practitioners and organizations responsible for such systems~\cite{SSAmodernization}.
As GenAI-based modernization approaches show promise but remain difficult to validate
and control, the need for representative evaluation resources becomes paramount.
Without corpus that reflect the structural and semantic characteristics of real production
COBOL, it is not possible to rigorously benchmark these approaches, characterize
their failure modes, or establish the conditions under which they can be trusted. 
%
% Scarcity of representative COBOL corpus
Representative COBOL corpus reflecting production mainframe programs are rarely
available in open access. Existing public collections are highly heterogeneous,
ranging from algorithmic exercises to domain-specific scripts, and do not exhibit
the structural patterns characteristic of enterprise legacy systems, such as
copybooks, sequential and indexed file handling, or business-oriented control
flows. The proprietary nature of enterprise COBOL codebases severely limits the availability
of realistic public corpus~\cite{lee2022}, and existing open-source collections
rely on filtering heuristics such as GitHub stars that do not reflect production
code quality~\cite{ali2023, dau2026}.

% Contribution
% To address these challenges, we propose a systematic agentic AI approach to the generation
% of realistic COBOL programs, combining agentic loop with syntactic validation. We
% further introduce an evaluation framework to assess the realism and structural complexity
% of generated programs, grounded in comparison with real production programs
% provided by an industrial partner and validated through expert review.

To address these challenges and the lack of strong benchmarks, we propose a systematic agentic AI approach to the
generation of realistic COBOL programs, combining agentic loop. The pipeline
first generates a syntactically valid and functionally correct program from a
natural language specification, then iteratively degrades it toward a calibrated
target complexity, using an asset library of patterns extracted from real
production code and complexity levels derived by correlation-based regression on
a reference corpus. 
Here, realism is understood as structural fidelity to production code, as captured by our metrics.

We evaluate whether iterative degradation reaches the target
complexity intervals derived from real production programs provided by our industrial
partner Sopra Steria, while business behavior is preserved in the process, and assess the practical
limits of the approach. We also  validate through expert review. Across three business
domains, our evaluation shows that the pipeline reliably produces syntactically
valid programs and moves them toward realistic structural complexity, while also
surfacing practical limitations, in particular around functional preservation
and scalability, which we discuss in detail. %\footnote{
%All COBOL programs generated and degraded by our approach, together with their input specifications, are

Our replication package is available as an artifact at
\textcolor{blue}{\url{https://github.com/jbespi/spec2cobolrot-samples.git}}.
%}

%% file: sections/related_work.tex
\section{Related Work}
\label{sec:related-work}

% Dataset translation
% COBOL-Coder -> ils font en plus de la recherche github, Java->COBOL avec GPT-4o
% Program augmentation
% [Saito et al., AISM 2025] augmentation de programmes existants pour entraîner des LLMs. C'est de l'augmentation, pas de la génération from scratch
% Full synthetic generation

Approaches to building COBOL datasets are centered on %corpora %for training and evaluating LLM-based
%modernization tools falls into three main categories: 
program translation,
program augmentation, and full synthetic generation. Mining public repositories
is a fourth, complementary source, but %as discussed in Section~\ref{sec:intro}, it
suffers from representativeness issues due to heterogeneous quality and GitHub-star-based
filtering heuristics~\cite{ali2023, lee2022}.

Program translation approaches construct COBOL corpora by translating programs from
a higher-resource source language. COBOL-Coder~\cite{dau2026} translates Java
programs into COBOL using an LLM, then validates the result through compiler-based
checks and back-translation similarity scoring. Cassano et al.~\cite{cassano2024knowledge} propose
MultiPL-T, which synthesizes and filters tests on high-resource language code
before translating it into low-resource languages, using test coverage as a validation
signal. These methods depend on the availability of a source-language corpus and
inherit its structural style. The resulting COBOL has no inherent reason to exhibit
the conventions of legacy production code, since it is derived from the syntax
and idioms of the source language rather than from COBOL itself.

Program augmentation approaches start from an existing, often small, code corpus
and transform it to increase diversity or realism. Saito et al.~\cite{saito2025}
propose grammar- and coverage-based augmentation, identifying underrepresented
grammar paths in a training corpus and prompting an LLM to generate code that
fills these gaps. Their approach targets syntactic diversity but not legacy-specific
structural realism. LintSeq~\cite{piterbarg2025} refactors real code into
sequences of linter-verified edits to increase solution diversity while
preserving functionality. Closest in spirit to our degradation phase, Leinonen
et al.~\cite{leinonen2024} prompt an LLM to imitate typical student mistakes and
show that the resulting error distribution is statistically close to that of real
student submissions. We pursue an analogous idea, injecting plausible legacy patterns
rather than plausible bugs, and targeting COBOL production realism rather than
student code.

Full synthetic generation approaches generate programs from scratch, typically via
instruction-based prompting, without transforming an existing corpus. Code Alpaca~\cite{codealpaca}
generates 20K synthetic instruction-solution pairs via Self-Instruct.
WizardCoder~\cite{luo2025} extends this with Evol-Instruct to iteratively
increase instruction complexity. 
Our approach combines elements of the latter two categories. Like full synthetic
generation, Phase~1 produces COBOL directly from a natural language
specification, without a source corpus to translate. Like program augmentation, Phase~2
transforms the generated program using patterns extracted from real production code.
Unlike prior augmentation work, this transformation targets a calibrated, corpus-derived
complexity profile rather than syntactic coverage or edit diversity alone.

%% file: sections/corpus_analysis.tex
\section{Corpus Analysis}

\subsection{Corpus Description}
\label{sec:corpus-description}

Our industrial partner provided us with a corpus of COBOL programs, which we used
to evaluate our work. This corpus of programs consist of 14 batch COBOL payroll
programs from a real legacy application. It covers several functions of a complete
processing pipeline, from the extraction and sorting of business files to
request generation, dictionary-based enrichment, result merging and payslip
production. From a technical perspective, the programs are characterized by a
file-oriented design, extensive use of copybooks and large global data structures.
The source code is incrementally developed and maintained over years. It's a representative
sample in terms of size, complexity of legacy COBOL programs.

\subsection{Structural Complexity Metrics}
\label{sec:metrics}

\begin{table}[htbp]
    \centering
    \caption{Structural complexity metrics.}
    \label{tab:metrics}
    \resizebox{\columnwidth}{!}{%
    \begin{tabular}{p{4.1cm}p{0.9cm}p{4.7cm}}
        \toprule \textbf{Metric}                                    & \textbf{Symbol} & \textbf{Rationale}                     \\
        \midrule \multicolumn{3}{l}{\textit{Program-level metrics}}  \\
        Source Lines of Code                                        & SLOC            & Program size excluding comments/blanks \\
        Number of paragraphs                                        & NPAR            & Structural decomposition granularity   \\
        Data/Procedure Div. LOC ratio                               & DIVR            & Balance between data and logic         \\
        Cyclomatic Complexity                                       & CC              & Control-flow complexity                \\
        Average paragraph length                                    & APL             & Code organization, modularity          \\
        \midrule \multicolumn{3}{l}{\textit{Legacy pattern metrics}} \\
        \midrule Number of GOTO statements                          & NGO             & Lack of structured control flow        \\
        Variable Usage Rate (\%)                                    & VUR             & Over-declaration                       \\
        Dead Paragraph Rate (\%)                                    & DPR             & Structural dead code                   \\
        \bottomrule
    \end{tabular}
    }
\end{table}

To assess the structural complexity of the COBOL programs, we selected a set of
metrics capturing scale, code complexity, maintenance drift, and modularization.
These combine classic metrics commonly used to characterize program-level
complexity with metrics specifically designed to capture traits typical of legacy
code, such as maintenance drift and control-flow degradation. Table~\ref{tab:metrics}
summarizes them.

For the scale dimension, we used Source Lines of Code (SLOC), number of paragraphs
(NPAR), and the Data/Procedure Division LOC ratio (DIVR). SLOC is a widely used
indicator of program size, and of the effort already invested and still required
to maintain it; it is important to keep this size realistic and not too small.
NPAR captures the structural decomposition of the code, i.e., the number of
minimal logical units in the program. DIVR assesses the balance between data and
logic: in legacy COBOL, the data division is often large due to extensive use of
global data structures and copybooks.

For the code complexity dimension, we used cyclomatic complexity (CC) and the
number of GOTO statements (NGO). CC measures the number of independent logic paths
through the code and is a good indicator of testing and maintenance difficulty; it
is typically high in legacy COBOL due to its procedural style with many
conditionals and loops. NGO captures the lack of structured control flow, often
associated with code that is hard to understand and maintain; combined with CC, it
gives a more complete picture of control-flow complexity, since CC alone is not
sensitive to control-flow structure.

For the maintenance drift dimension, we used the Variable Usage Rate (VUR) and the
Dead Paragraph Rate (DPR). VUR measures the proportion of declared variables never
referenced in the procedure division, capturing variable over-declaration, common
in legacy COBOL. DPR measures the proportion of paragraphs never executed. Together,
these metrics assess how far the program has drifted from its original design, and
how much noise this drift adds for modernization tools relying on generative AI.

For the modularization dimension, we used the Average Paragraph Length (APL), the
mean number of statements per paragraph. Very short paragraphs fragment related
logic across units, while very long ones concentrate too much responsibility in a
single one, increasing the difficulty for generative AI tools during modernization.

\subsection{Descriptive Statistics}
\label{sec:descriptive-stats}

Structural metrics are computed using two tools provided by our industrial partner:
a static extractor that parses each COBOL program to extract structural and
syntactic information, and an analyzer that computes the metrics of Table~\ref{tab:metrics}
from this extracted information. Table~\ref{tab:metric-stats} reports summary
statistics (minimum, maximum, mean, median, and standard deviation) for each
structural metrics across the 14-program corpus. Given the limited size of this corpus and the specific scope,
these statistics should be interpreted as descriptive characteristics rather
than generalizable estimates of COBOL legacy systems.

\begin{figure*}[ht]
    \centering
    \includegraphics[width=0.9\textwidth]{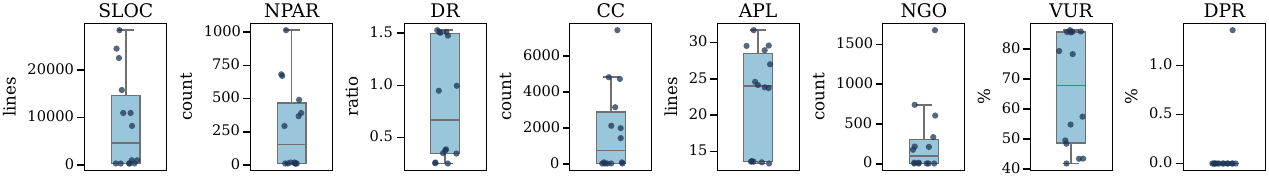}
    \Description{Eight box plots showing the distribution of structural complexity metrics (SLOC, NPAR, DIVR, CC, APL, NGO, VUR, DPR) across the 14-program COBOL corpus, with individual program values overlaid as points. Most metrics show right-skewed distributions with a majority of programs at moderate values and a few outlier programs reaching much higher complexity.}
    \caption{%Distribution of structural complexity 
    Metrics across the 14-program
    corpus. Boxes show interquartile ranges; individual points represent each
    program.}
    \label{fig:corpus_metric_distributions}
\end{figure*}

Figure~\ref{fig:corpus_metric_distributions} complements Table ~\ref{tab:metric-stats}
by visualizing the shape of each metric's distribution, overlaying individual program
values. Most metrics exhibit dispersion and right-skewed distributions, with a majority
of programs concentrated at moderate values and a small number of outlier programs
reaching high complexity. This is particularly visible for SLOC, CC and NGO,
where a single program (the same across all three metrics) exhibits values
several times higher than the corpus median, indicating some degraded legacy implementation
coexisting with comparatively cleaner programs in the same industrial codebase.

In contrast, DPR shows nearly all programs with a value of 0\% (median = 0\%), with
the exception of the same outlier program reaching 1.36\%. This suggests that dead
paragraphs are not a widespread issue in this corpus, but can be significant in
modernization use cases. That's why we keep this metric here in our corpus analysis.

\begin{table}[t]
    \centering
    \caption{Structural metric distributions on the 14-programs} % corpus (Sopra     Steria).}
    \label{tab:metric-stats}
    \scalebox{0.8}{
    \begin{tabular}{c r r r r r c}
        \toprule \textbf{Symbol} & \textbf{Min} & \textbf{Max} & \textbf{Mean} & \textbf{Median} & \textbf{Std} & \textbf{Unit} \\
        \midrule SLOC            & 360          & 28387        & 8957.14       & 4660.0          & 10203.45     & lines         \\
        NPAR                     & 12           & 1015         & 286.71        & 157.0           & 329.83       & count         \\
        DIVR                     & 0.25         & 1.53         & 0.83          & 0.67            & 0.57         & ratio         \\
        CC                       & 29           & 7434         & 1859.43       & 765.5           & 2369.94      & count         \\
        APL                      & 13.33        & 31.72        & 22.20         & 24.00           & 7.12         & lines         \\
        \midrule NGO             & 4            & 1678         & 286.71        & 95.5            & 465.52       & count         \\
        VUR                      & 41.893       & 86.316       & 66.163        & 67.925          & 18.947       & \%            \\
        DPR                      & 0.00         & 1.36         & 0.10          & 0.00            & 0.36         & \%            \\
        \bottomrule
    \end{tabular}
    }
\end{table}

\subsection{Inter-Metric Correlation}
\label{sec:correlation}

To assess if our metrics vary independently or jointly, we compute pairwise
Spearman correlations across the corpus. Size and control-flow-related metrics (SLOC,
NPAR, CC, NGO, VUR) are strongly correlated with each other ($\rho > 0.90$),
while DR is strongly anti-correlated with this group ($\rho < -0.90$), and APL shows
only moderate correlation ($\rho \approx 0.6$--$0.7$). DPR remains largely independent
of all other metrics ($|\rho| < 0.25$). These %results 
indicate that complexity
manifests as a coherent, multi-dimensional pattern rather than as independently varying
properties, motivating the correlation-based approach for defining target
complexity levels (see Section~\ref{sec:exp-setup}).

%% file: sections/approach.tex
\section{Approach}

%\subsection{Overview}

\begin{figure*}[htbp]
    \centering
    \includegraphics[width=0.8\textwidth]{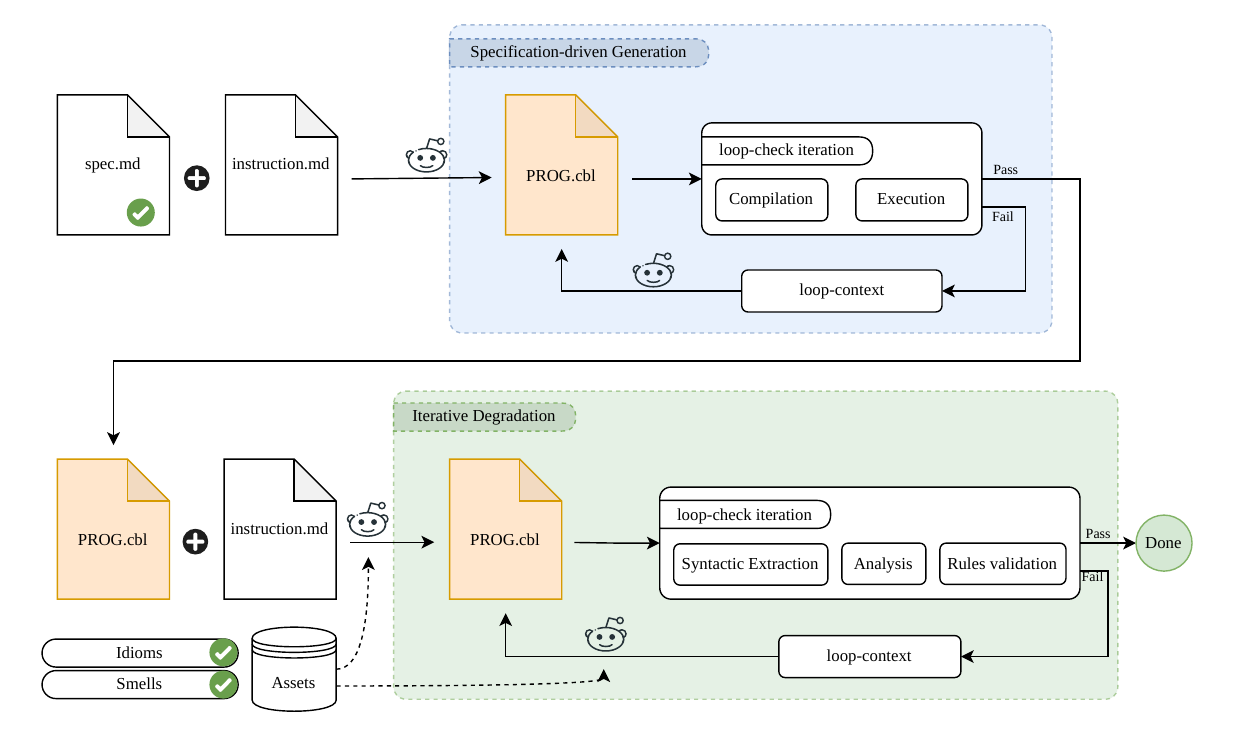}
    \Description{Flowchart of the Spec2COBOLRot pipeline in two phases. Top, Specification-driven Generation: a spec.md and instruction.md feed a generation agent producing PROG.cbl, which enters a compilation and execution loop; on failure the loop-context feeds back for another iteration, on pass the program moves to the next phase. Down, Iterative Degradation: an asset library of idioms and smells, combined with syntactic extraction, analysis, and rules validation, guides an agent that modifies PROG.cbl using instruction.md, inside a loop-check iteration with loop-context feedback on failure, until the process reaches Done on pass.}
    \caption{Overview of the Spec2COBOLRot pipeline.}
    \label{fig:pipeline}
\end{figure*}

Figure~\ref{fig:pipeline} illustrates the overall pipeline of our approach. The
pipeline consists of two main phases: Specification driven COBOL Generation and Iterative
degradation. The first phase focuses on generating COBOL programs based on
provided specifications, while the second phase iteratively degrades these programs
to enhance their realism by incorporating patterns from an asset library. This
modular approach allows for efficient handling of multiple specifications and
facilitates the integration of various components in the pipeline.

We %deliberately 
separate generation from degradation rather than producing a
degraded program directly. A program already containing legacy patterns is difficult
for a domain expert to read and validate against the specification. By first
producing a clean program, we obtain a version that can be readily reviewed against
the specification, and that serves as the reference against which the degraded
program's input/output behavior is later compared (Section~\ref{sec:rq2}). This separation
is what allows us to certify that a realistic degraded program still conforms to
its original specification, a guarantee that is particularly valuable for
modernization use cases such as business rule extraction or retro-documentation,
where the generated benchmark's ground truth must be known with confidence.

\subsection{Inputs: Specifications and Asset Library}
\label{sec:approach-spec-asset}

\textit{\textbf{Specifications.}} Each task in the pipeline is driven by a natural
language specification that describes the business logic and requirements of a target
COBOL program. Specifications are generated from real-world use cases with an
agent and subsequently validated by a human expert from our industrial partner
to ensure alignment with real enterprise use cases. Specifications are provided in
a structured format that captures the essential elements of the program's functionality,
including functional objective, business context, required inputs, produced outputs,
and business rules.

The specifications of our running examples, described in
Table~\ref{tab:specs}, are available in our replication package. Each
specification comprises four levels: the functional objective, an input
data definition, an output data definition, and a business rule expressed
as an explicit computation.

\textit{\textbf{Asset Library.}} To guide the degradation phase, we rely on an
asset library containing recurring patterns extracted from real production COBOL
programs. Each asset belongs to one of three families: \textbf{idioms} are recurring
code fragments characteristic of legacy COBOL batch programs, reflecting common
operational conventions rather than any defect; \textbf{smells} are poor coding
practices that erode maintainability, such as unstructured control flow or dead state,
without altering functional behavior; \textbf{mismatches} are COBOL-specific
constructs that force a non-trivial adaptation decision during later
modernization. Each asset is validated to ensure it accurately reflects a pattern
observed in real-world code, so the library serves as a grounded reference for
the degradation agent rather than an arbitrary source of complexity.

Beyond its family, each asset is described by a short rationale,
a canonical code form, and guidance on when and how it should
be injected. The full set of assets can be found in our replication package (the full library
composition is reported in Section~5).

\subsection{Phase 1: Specification driven COBOL Generation}
\label{sec:spec-driven-generation}

\textit{\textbf{Generation.}} Given a specification and an instruction prompt \linebreak (\texttt{instruction.md}),
the generation agent is responsible for producing an initial COBOL program that
adheres to the specified requirements. Data files are generated alongside the COBOL
program to ensure that the program can be compiled and executed successfully.
Synthetic data is generated from the specification context. This synthetic data is
reused in Phase 2 to verify that degradation preserves the program's input/output
behavior (Section~\ref{sec:degradation-phase}).

\textit{\textbf{Syntactic validation.}} The generated COBOL program undergoes
syntactic validation through an iterative loop. At each iteration, the program
is compiled using GnuCOBOL and executed against the generated data files. If compilation
and execution both succeed, the program is passed to Phase 2. Otherwise, the compiler
and runtime errors are fed back to the generation agent as fresh context (loop-context,
Figure~\ref{fig:pipeline}), which produces a revised program for the next iteration.
This loop is bounded by a budget of $k$ iterations. This step is crucial to confirm
that the generated code is not only syntactically correct but also executable,
providing a foundation for degradation in the next phase.

\textit{\textbf{Output.}} %The output of Phase 1 
It is a syntactically correct and executable
COBOL program that meets the requirements specified in the input specification.
However, at this stage, the program may lack the realism and complexity observed
in production COBOL code, which will be addressed in Phase 2 through iterative degradation.
If the program still fails to compile or execute after $k$ iterations, the pipeline
terminates for that specification and Phase 2 is not launched.

\subsection{Phase 2: Iterative Degradation}
\label{sec:degradation-phase}

\textit{\textbf{Asset injection.}} An degradation agent is employed to iteratively
enhance the realism of the generated COBOL program by incorporating patterns from
the asset library. With an other instruction prompt (\texttt{instruction.md}),
the agent analyzes the initial program and identifies opportunities to inject
relevant assets that align with the program's structure and functionality. This process
involves selecting appropriate patterns from the library and integrating them into
the codebase, thereby increasing its complexity and resemblance to real-world
COBOL programs. Assets are carefully chosen to ensure that they do not violate the
original specification and maintain the program's correctness while enhancing its
realism.

\textit{\textbf{Deriving Complexity Levels.}} To evaluate the pipeline's ability
to generate realistic programs across the complexity spectrum, rather than at isolated
points, we define target complexity levels as intervals over SLOC, spanning a reference
corpus's observed range. For each level, we derive a target interval for metrics
strongly correlated with SLOC ($R^{2}> 0.7$). With linear regression, we predict
the metric at both interval bounds, take the envelope of the two predictions, and
widen it by $\pm 1$ residual standard deviation to account for the variability not
explained by SLOC alone. Lower bounds are clipped at the corpus-wide minimum
observed value for each metric, avoiding physically implausible targets that could
otherwise result from residual extrapolation at high SLOC values. Metrics whose
correlation with SLOC is too weak to support a reliable prediction are instead
assigned the corpus-wide observed range. This method requires only the structural
metrics already computed on the reference corpus (Section~\ref{sec:descriptive-stats}),
and can be applied to any corpus of production programs to derive realistic, complexity-scaled
degradation targets.

\textit{\textbf{Stopping criterion.}} The target intervals derived above are
used to drive and stop the degradation loop. They are computed once, ahead of
the loop, by using the \textit{Deriving Complexity Levels} method. In our experiments,
these are the min--max bounds derived per complexity level (Section~\ref{sec:exp-setup}).
At each iteration, the structural metrics defined in Section~\ref{sec:metrics} (Table~\ref{tab:metrics})
are recomputed on the \emph{current} program using the extraction and analysis
tools. If any metric falls outside its target interval, degradation continues
with further asset injection; otherwise, the process stops. If the target intervals
are not reached within a budget of $k$ iterations, the process stops after the
last iteration and returns the program in its current state. The pipeline does not
depend on target intervals being derived from a real reference corpus. In the
absence of such a corpus, these bounds could instead be set manually or from other
sources, at the risk of driving degradation toward target profiles that are less
representative of actual production code.

%% file: sections/experimental_setup.tex
\section{Experimental Setup}
\label{sec:exp-setup}

% explication des 3 specs utilisés briévement
% design experiement 3 specs x 3 complexity level = 9 programs generated
% outil de mesure GnuCOBOL, cobol-extractor, cobol-analyzer
% Protocole de revue expert ?

\begin{table}[htbp]
    \centering
    \caption{Specifications used in the evaluation.}
    \label{tab:specs} \small
    
    \begin{tabular}{@{}lp{0.62\linewidth}@{}}
        \toprule Name     & Description                                                                                                                                                       \\
        \midrule TRNRECBT & Nightly banking transaction processing against account positions, with fee and interest application and end-of-day interbank reconciliation.                      \\
        PAYCOMBT          & Monthly payroll computation with social contributions, tax withholding, and ordered deductions, producing pay slips, bank transfers, and a statutory declaration. \\
        CLMPREMT          & Annual insurance premium recomputation with experience-based adjustment, and claim settlement across motor, property, and health portfolios.                      \\
        \bottomrule
    \end{tabular}
\end{table}

\textit{\textbf{Specifications Used.}} We use three specifications from three
business domains (banking, payroll, insurance), each generated from a real-world
use case and validated by a COBOL expert (Table~\ref{tab:specs}). 
% PAYCOMBT is used as the running example throughout Section~\ref{sec:approach-spec-asset}. 

\textit{\textbf{Asset Library Construction.}} \label{sec:asset-library-setup} The
library was created and extracted from the production programs provided by our industrial
partner (Section~\ref{sec:corpus-description}) and contains 26 assets: 8 idioms,
15 smells, and 3 mismatches (Table~\ref{tab:asset-library}).

\begin{table}[htbp]
    \centering
    \caption{Asset library composition.}
    \label{tab:asset-library} \small
    \begin{tabular}{@{}lrp{0.55\linewidth}@{}}
        \toprule Family & \# & Examples                                                                               \\
        \midrule Idioms & 8  & EOF-driven iteration; indexed lookup; reject-record output                             \\
        Smells          & 15 & unstructured GO TO flow; working-storage overdeclaration; mixed I/O and business logic \\
        Mismatches      & 3  & memory-overlay views; packed-decimal representation, redefine                          \\
        \bottomrule
    \end{tabular}
\end{table}

\textit{\textbf{Tools and configuration.}} The generation and degradation agents
are both using GPT-5.4 . Each phase is bounded by an iteration budget of $k=5$
loop iterations (Figure~\ref{fig:pipeline}): Phase 1 recompiles and re-executes the
program up to 5 times to reach a syntactically valid and executable state; Phase
2 performs up to 5 asset-injection iterations to reach the target complexity
profile. % (Section~\ref{sec:}). 
Syntactic validation relies on GnuCOBOL;
structural metrics are computed using the extraction and analysis tools provided
by our industrial partner. The pipeline is orchestrated with Harbor~\cite{Harbor_Framework}.

\textit{\textbf{Target Complexity Levels.}} The four complexity levels are
defined by manually chosen SLOC intervals, spanning the observed range of the industrial
partner corpus (Section~\ref{sec:descriptive-stats}, Table~\ref{tab:metric-stats})
from small ($\sim$2.5k SLOC) to large ($\sim$25k SLOC) programs. These SLOC
intervals are the only manually set parameter of the method; all other target intervals
are derived from them via regression. Applying the \textit{Deriving Complexity
Levels} method of Section~\ref{sec:degradation-phase} to the industrial partner corpus,
NPAR, DIVR, CC, NGO, and VUR are strongly correlated with SLOC ($R^{2}> 0.7$)
and are assigned regression-based intervals; APL's correlation is too weak ($R^{2}
= 0.36$) and is instead assigned the corpus-wide observed range. DPR is excluded
due to its near-zero variance across the corpus.

Table~\ref{tab:levels} reports the resulting target intervals for four
complexity levels. Note that intervals for adjacent levels may partially overlap,
reflecting residual variability around the SLOC-metric trend rather than an
inconsistency in the level definitions.

%% file: sections/evaluation.tex
\section{Evaluation}
% final score of programs on metrics (are they in ranges of real programs)
% comparing before/after enrichment phase *
% expert review (qualitative independant judgment)
To evaluate the effectiveness of our agentic pipeline, we designed a set of research
questions that focus on the syntactic validity, structural completeness, and
realism of the generated COBOL programs. 
%The evaluation is structured around
%three main RQs:

\begin{description}
     \item[\textbf{RQ1}] Does the agentic pipeline generate syntactically valid and
        structurally complete COBOL programs from natural language specifications
        ?
    % compilation with GnuCOBOL on generated programs, how many programs compile ? all ?
    % presence of 4 divisions ?
    % execution successful on test data
    % data and business rules respected ?

    \item[\textbf{RQ2}] Does the iterative degradation process increase the target
        structural complexity while preserving their functional behavior ?

    % metrics before/after degradation and at each iteration vs target metrics
    % how many iterations needed to reach target metrics ? correlation between number of iterations and distance to target metrics ?
    % diff I/O before/after degradation...

    \item[\textbf{RQ3}] What are the practical scale limits beyond which the degradation process fails to produce compilable and functionnaly equivalent COBOL programs ?
    % success/failure per complexity level, accross programs
    % which metric(s) most often cause compilation failure or behavioral divergence ?
    % observation on the degradation process (which metrics are more difficult to reach, which are easier, ...

    \item[\textbf{RQ4}] Do domain experts perceive the generated programs as realistic
        production COBOL ?

    % review protocol (experts outside of this work, how many experts...)
    % qualitative review of generated programs ...
    % feedback on the coherence from their experience on real COBOL programs...
\end{description}

\begin{table}[htbp]
    \centering
    \caption{Correlation of target intervals for our metrics across
 complexity levels, derived by linear regression on SLOC.}
    \label{tab:levels} \footnotesize
    \begin{tabular}{l c c c c}
        \toprule \textbf{Metric} & \textbf{L1 (2.5--3.5k)} & \textbf{L2 (4.5--6.5k)} & \textbf{L3 (8--12k)} & \textbf{L4 (20--25k)} \\
        \midrule NPAR            & 33--159                 & 98--255                 & 210--431             & 594--847              \\
        DIVR                     & 0.80--1.44              & 0.66--1.34              & 0.39--1.18           & 0.25--0.61            \\
        CC                       & 29--1009                & 439--1696               & 1240--2955           & 3988--5932            \\
        NGO                      & 4--293                  & 4--413                  & 24--633              & 504--1153             \\
        APL                      & 13.3--31.7              & 13.3--31.7              & 13.3--31.7           & 13.3--31.7            \\
        VUR (\%)                 & 68.6--84.3              & 63.4--80.8              & 53.9--74.8           & 41.9--54.1            \\
        \bottomrule
    \end{tabular}
\end{table}

\subsection{RQ1 - Syntactic Validity}
\label{sec:rq1}

For each generated program, we verify that it compiles with GnuCOBOL together and
executes with data files without runtime errors, and produces all input and
output files required by the specification. Across all generation runs for the three
specifications (TRNRECBT, PAYCOMBT, CLMPREMT), every program satisfies all three
checks. 
Beyond these automated checks, manual inspection of the generated code against the
specification suggests that the key business rules, in particular the
computations explicitly required by the specification, are mostly reflected in the
procedure division. Data formats specified for input and output fields
(types, lengths, value ranges) are also respected in most cases. %the large majority of cases.
This inspection is qualitative and not exhaustive; a systematic, rule-by-rule
and field-by-field validation is left for future work.

\begin{comment}

\begin{tcolorbox}
    [ colback=blue!10, colframe=blue!50, boxrule=0.5pt, arc=4pt, left=6pt, right=6pt,
    top=4pt, bottom=4pt ]

    \textbf{Summary.} The agentic pipeline reliably generates syntactically valid
    and structurally complete COBOL programs from natural language specifications,
    with generated code consistently reflecting the computations and data
    formats required by the specification.
\end{tcolorbox}

\end{comment}

\subsection{RQ2 - Realism Degradation}
\label{sec:rq2}

% IL FAUDRA AFFICHER AVG sur 3 attempts
% In order to evaluate the utility of the degradation process, we run 3 attempts on
% each spec, in order to take into account variability. So values in diagram are averages
% of 3 attempts per program.

\begin{figure*}[htbp]
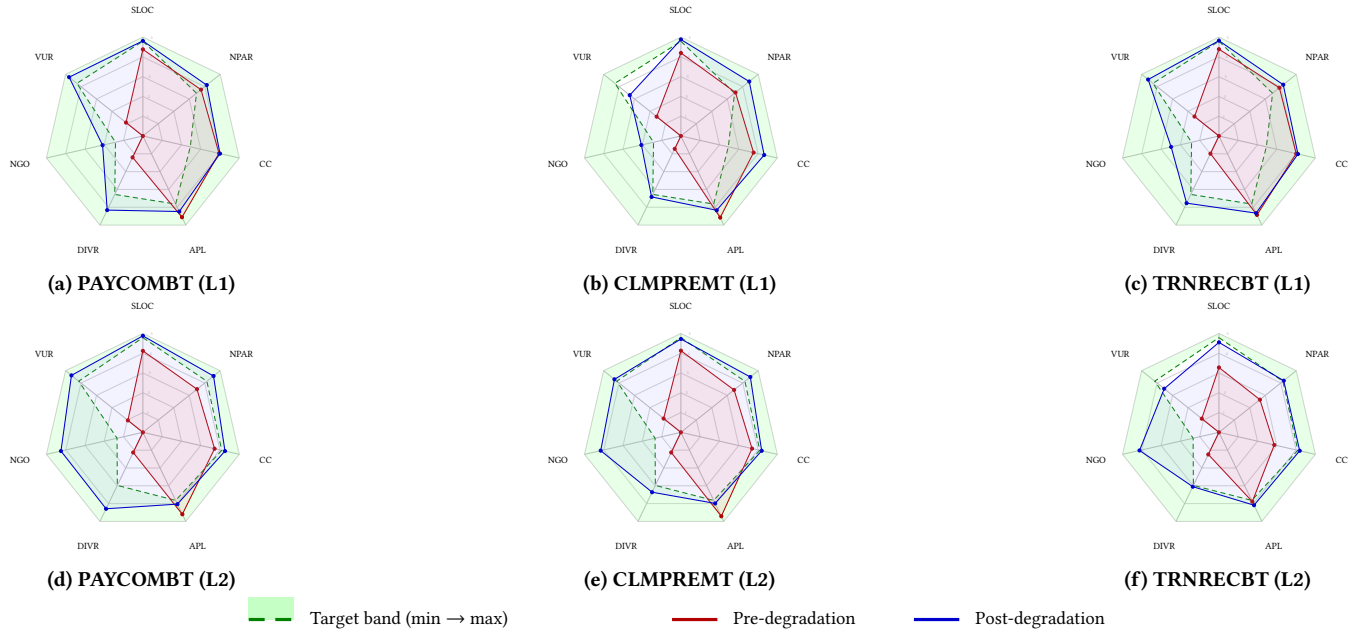

    \centering
    \Description{Six radar charts arranged in two rows of three, comparing structural complexity metrics (SLOC, NPAR, CC, APL, DIVR, NGO, VUR) before and after degradation. Top row shows PAYCOMBT, CLMPREMT, and TRNRECBT at complexity level L1; bottom row shows the same three programs at level L2. Each chart displays a target band, a pre-degradation profile mostly inside the band, and a post-degradation profile expanding outward toward or into the band on most axes.}
    % --- Ligne L1 ---
    \begin{subfigure}[t]{0.2\textwidth}
        \centering
        \resizebox{\linewidth}{!}{%
        \KiviatDiagram {4.79, 3.47, 2.46, 3.82, 3.28, 1.42, 4.27} % min
        {4.38, 3.76, 3.95, 4.56, 1.2, 0.0, 1.09} % init
        {4.81, 4.13, 4.00, 4.24, 4.16, 2.09, 4.77} % final
        }
        \caption{PAYCOMBT (L1)}
    \end{subfigure}
    \hfill
    \begin{subfigure}[t]{0.2\textwidth}
        \centering
        \resizebox{\linewidth}{!}{%
        \KiviatDiagram {4.79, 3.47, 2.46, 3.82, 3.28, 1.42, 4.27} % min
        {4.2, 3.53, 3.77, 4.58, 0.72, 0.0, 1.57} % init
        {4.88, 4.42, 4.32, 4.17, 3.42, 2.05, 3.31} % final
        }
        \caption{CLMPREMT (L1)}
    \end{subfigure}
    \hfill
    \begin{subfigure}[t]{0.2\textwidth}
        \centering
        \resizebox{\linewidth}{!}{%
        \KiviatDiagram {4.79, 3.47, 2.46, 3.82, 3.28, 1.42, 4.27} % min
        {4.39, 3.91, 4.01, 4.43, 0.99, 0.0, 1.58} % init
        {4.82, 4.16, 4.1, 4.33, 3.77, 2.48, 4.58} % final
        }
        \caption{TRNRECBT (L1)}
    \end{subfigure}

    % --- Ligne L2 ---
    \begin{subfigure}[t]{0.2\textwidth}
        \centering
        \resizebox{\linewidth}{!}{%
        \KiviatDiagram {4.79, 4.14, 4.09, 3.82, 2.98, 1.34, 4.15} % min
        {4.12, 3.5, 3.72, 4.6, 1.13, 0.0, 0.97} % init
        {4.88, 4.57, 4.26, 4.03, 4.29, 4.25, 4.62} % final
        }
        \caption{PAYCOMBT (L2)}
    \end{subfigure}
    \hfill
    \begin{subfigure}[t]{0.2\textwidth}
        \centering
        \resizebox{\linewidth}{!}{%
        \KiviatDiagram {4.79, 4.14, 4.09, 3.82, 2.98, 1.34, 4.15} % min
        {4.13, 3.44, 3.69, 4.71, 1.13, 0.0, 1.12} % init
        {4.72, 4.49, 4.19, 3.99, 3.36, 4.16, 4.3} % final
        }
        \caption{CLMPREMT (L2)}
    \end{subfigure}
    \hfill
    \begin{subfigure}[t]{0.2\textwidth}
        \centering
        \resizebox{\linewidth}{!}{%
        \KiviatDiagram {4.79, 4.14, 4.09, 3.82, 2.98, 1.34, 4.15} % min
        {3.28, 2.65, 2.87, 3.89, 1.24, 0.0, 1.11} % init
        {4.55, 4.19, 4.19, 4.09, 3.05, 4.12, 3.54} % final
        }
        \caption{TRNRECBT (L2)}
    \end{subfigure}

    \KiviatLegend
    \caption{Comparing the realism of generated COBOL before/after the degradation phase, for the L1 (top row) and L2 (bottom row) target complexity levels.}
    \label{fig:radar-L1-L2}
\end{figure*}

%We evaluate RQ2 at the lowest complexity level L1~\ref{sec:exp-setup}. %Scale effects at higher complexity levels are addressed separately in RQ3 (Section~\ref{sec:rq3}).
For each of the three specifications (TRNRECBT, PAYCOMBT, CLMPREMT), we run the
pipeline 3 times and report results per program. 
Structural complexity increases across all three programs after degradation. Figure~\ref{fig:radar-L1-L2}
reports, for each program, the structural metrics before (pre-degradation, red)
and after (post-degradation, blue) the degradation phase, together with the
calibrated target band for L1 and L2 levels. %(green, Section~\ref{sec:levels}). 
Each axis is log\textsubscript{10}-scaled
and normalized so that a raw value of zero maps to the center and the upper
bound of the target interval maps to the outer edge, making relative progress toward
the target comparable across metrics of very different raw scale (SLOC, CC, and
NGO span several orders of magnitude in the reference corpus~\ref{sec:descriptive-stats}).
Across all three programs, the pre-degradation profile lies inside the target band
on most axes, reflecting the intentionally simple programs produced by Phase~1.
After degradation, the profile expands outward and enters the target band on nearly
every axis, indicating that asset injection systematically moves the program
toward the complexity levels observed in the reference corpus.

Functional equivalence, however, is not always preserved. We check functional equivalence
by executing both the pre- and post-degradation program on the same synthetic
test dataset (Section~\ref{sec:spec-driven-generation}) and diffing their output
files. 
%%% ajouter 
Note that non-terminating programs are caught by an execution timeout and rejected earlier in the loop.
Results on L1 are mixed: all 3 runs of CLMPREMT preserve output files unchanged,
while 2 of 3 runs of PAYCOMBT and 2 of 3 runs of TRNRECBT introduce differences across
all output files, for an overall preservation rate of 5/9 runs. 
This choice reflects
a deliberate design trade-off, we favored giving the degradation
agent freedom to explore structural and pattern-level transformations without an
additional hard constraint at every iteration, to avoid over-restricting its
ability to reach the target complexity profile. In hindsight, our results
suggest this freedom comes at a measurable cost to functional preservation, and that
a tighter, per-iteration verification loop, similar to the one already used in Phase~1
(Section~\ref{sec:spec-driven-generation}), would likely be a better trade-off
in practice (Section~\ref{sec:discussion}).

\begin{comment}

\begin{tcolorbox}
    [ colback=blue!10, colframe=blue!50, boxrule=0.5pt, arc=4pt, left=6pt, right=6pt,
    top=4pt, bottom=4pt ]

    \textbf{Summary.} The degradation process reliably increases structural complexity
    toward the calibrated target across all tested programs, but functional
    equivalence is preserved in only 5 of 9 runs, showing that structural realism
    and functional preservation are not jointly guaranteed by the current design.
\end{tcolorbox}

\end{comment}

\subsection{RQ3 - Practical Scale Limits}
\label{sec:rq3}

Across all generated programs for L3 and L4, we did not observe practical
scale limits in terms of compilability or functional equivalence: every
degraded program compiled successfully, and the data manipulated between
the pre-degradation and post-degradation phases remained often consistent. The
practical limit we identified is instead qualitative in nature and relates
to realism rather than correctness. When the degradation process is
constrained to reach a target size or complexity threshold, the underlying
LLM tends to compensate by duplicating paragraphs that are structurally
and semantically very similar to one another, rather than introducing
genuinely diverse legacy-like constructs. While this strategy allows the
imposed thresholds to be met, it comes at the cost of realism, as such
repetitive patterns are not representative of how legacy COBOL programs
typically grow in size and complexity over time. 
This suggests that scale constraints should be interpreted not as a limit
on the technical feasibility of the degradation process, but rather as a
factor that can silently erode the realism of the generated defects.

\subsection{RQ4 - Expert Review}

To validate these findings, the generated specifications were reviewed by 3
domain experts with 20 to 30 years of experience in legacy systems, and the
resulting programs by 2 other experts with comparable experience. Overall,
experts were impressed by the agent's ability to generate COBOL code for a
complex business domain from the specifications, though the resulting programs
were judged very clean, arguably too clean for real industrial COBOL. Figure~\ref{fig:expert-realistic-excerpt}
shows one excerpt from PAYCOMBT that experts recognized as a typical legacy
pattern encountered during modernization audits. Other structural aspects,
however, revealed the limits of the injected patterns: the deliberate absence of
\texttt{COPY} statements, ubiquitous in real legacy codebases; \texttt{IF} blocks
systematically closed with an explicit \texttt{END-IF} rather than the more
common, error-prone implicit termination by period (\texttt{.}); an overuse of
\texttt{STRING} statements with \texttt{DELIMITED BY SIZE}, several also relying
recurrently on \texttt{|} as a delimiter; and \texttt{REDEFINES} clauses, which
our degradation process rarely managed to introduce due to the lack of
sub-variable format definitions in the generated data structures. Overall,
experts considered the use of legacy assets a promising first step, but still too
synthetic to fully capture the diversity of real legacy defects.

\begin{figure}[htbp]
    \centering
    \begin{tcolorbox}
        [ colback=gray!5, colframe=gray!60, boxrule=0.4pt, arc=1pt, left=6pt, right=6pt,
        top=5pt, bottom=5pt, fontupper=\small ]

        \hspace{10pt}
        \begin{minipage}{0.85\linewidth}
            \begin{verbatim}
01  WS-AMTS.
    05  WS-ADJ-SAL          PIC 9(7)V99.
    05  WS-HR-RATE          PIC 9(7)V99.
    05  WS-OT1-AMT          PIC 9(7)V99.
...
COMPUTE WS-HR-RATE =
    WS-ADJ-SAL / CAL-HRS (WS-CAL-IDX)
COMPUTE WS-OT1-AMT =
    AGG-OT1 (WS-I) * WS-HR-RATE * 1.25
\end{verbatim}
        \end{minipage}
    \end{tcolorbox}
    \Description{COBOL code excerpt in a gray bordered box, showing a WS-AMTS data structure with three PIC 9(7)V99 fields (WS-ADJ-SAL, WS-HR-RATE, WS-OT1-AMT), followed by two COMPUTE statements deriving WS-HR-RATE from WS-ADJ-SAL divided by CAL-HRS, and WS-OT1-AMT from AGG-OT1 multiplied by WS-HR-RATE and a factor of 1.25.}
    \caption{Excerpt from the degraded PAYCOMBT program}
    \label{fig:expert-realistic-excerpt}
\end{figure}

%% file: sections/threats.tex
\section{Threats to Validity}
% construct validity:
% complexity/quality metrics are syntactic indicators and dot not directly measure business domain comprehension difficulties. This is mitigated by the fact that experts validated the specifications documents as something realistic.
% limited number of generated programs (construct validity)
% ranges calibrated on 15 programs from a single industrial partner Sopra Steria on one business domain (hr payroll)

% expert subjectivity (internal validity)

% LLM exact reproducibility (external validity)

\textbf{Construct validity.} The structural metrics used in this study are based on
syntactic extraction that do not directly measure the comprehension difficulties
of the business domain. This threat is mitigated by the fact that the
specification documents were validated by experts as realistic, although expert
review remains subjective and may vary across reviewers with different experience.
Moreover, we deliberately targeted a level of structural complexity comparable to production programs, which increases the representativeness of our
generated programs relative to real-world code and contributes to the overall
realism of the study.

\textbf{Internal validity.} The generation and enrichment steps rely on non-deterministic
LLM calls. Running the pipeline on the same specification may produce different
programs, this limits exact reproducibility. So the overall pipeline is not
expected to be exactly reproducible. But it's mitigated by the fact that in our use
case, we want variability in the generated programs to cover a wide range of
possible implementations.

\textbf{External validity.} Our study relies on a limited number of generated programs, though spanning distinct domains (banking, payroll, insurance), which constrains the generalization of our results. The reference metrics are derived from 14 programs provided by a single industrial partner in one business domain (HR payroll), which may further limit generalization to other enterprises and domains. Our pipeline is nonetheless designed to be generic, and applicable to arbitrary business contexts, domains, and COBOL dialects, and could be extended to cover a wider range of PIC clause variations and more complex \texttt{DATA DIVISION} structures. Executing programs relying on such constructs, or on CICS and embedded SQL, is not a limitation of the methodology, but requires an execution environment capable of supporting them, which our current toolchain does not provide.

%% file: sections/discussion.tex
\section{Discussion}
\label{sec:discussion}

% what the iteractive loop actually delivers
% limit on this approach
% perspectives : scale up ? database ? other business domains ?
% croiser les résultats entre RQ ?  est-ce que les métriques difficiles en RQ2/RQ3 sont aussi celles jugées "peu réalistes" par les experts en RQ4 ?

%The degradation process successfully increases the
%structural complexity of the generated programs towards the reference corpus.
%Functional correctness is verified by input/output comparison between original and
%degraded programs on a synthetic test dataset, confirming identical outputs on the
%same inputs. This guarantee is nonetheless limited by test data coverage. Future
%work could account for this coverage and investigate the impact of degradation on
%test coverage and runtime performance.

The degradation process successfully increases the structural complexity of the
generated programs towards the reference corpus. Functional correctness, verified
by input/output comparison on a synthetic dataset, holds only in part and remains
limited by test data coverage, an impact future work could also extend to
runtime performance.

Structural complexity is primarily driven by business logic and shaped by
decades of patches and modifications, so targeting absolute metric values
independently of business complexity may appear artificial. This does not
undermine our objective: our goal is to produce programs exhibiting the
structural characteristics of legacy code (poor practices, accumulated
complexity, maintenance drift) while remaining functionally correct. A program
"too complex" for its business logic is an even harder case for modernization
tools, and therefore a more valuable benchmark artifact.
%%% comparison with LLM
 A natural baseline is direct LLM degradation from a generic "make it look like legacy code" prompt, without the asset library. 
We made the opposite choice to keep injected patterns traceable to real production code rather than the model's priors, and leave a quantitative comparison of the two to future work.

A promising direction is to keep the decoupling between generation and
degradation, but replace the degradation process itself with a more realistic
alternative: rather than injecting patterns directly, a maintenance plan
(a to-do list of patches and modifications spanning several decades) would drive a
legacy-developer agent that incrementally adds and modifies code as would have
happened over time. This more realistic process could
plausibly yield more realistic programs, though at a higher generation cost, and
we leave its evaluation to future work.

We did not include comment-related metrics, such as comment density or
comment-code alignment. The latter, capturing whether comments remain
semantically consistent with the code, would be useful for assessing
documentation reliability in legacy systems, where documentation is often
outdated~\cite{sabetto2025comments}. However, it requires semantic analysis
beyond static structural parsing, which we leave for future work.
% parler d'ajouter un check diff sur les données dans la loop de degradation.

%% file: sections/conclusion.tex
\section{Conclusion}

We proposed a systematic agentic AI pipeline for generating realistic COBOL
programs, combining specification-driven generation with iterative degradation
guided by patterns and complexity targets extracted from real production code.
%Our evaluation across three programs from distinct business domains showed that
%degradation reliably produces syntactically valid programs and reaches target
%complexity levels while preserving business behavior, yielding programs more
%representative of real-world systems. 
Our evaluation across three programs from distinct business domains showed that degradation reliably produces syntactically valid programs and moves them to the target complexity levels, with business-behavior preservation achieved in part, yielding programs that exhibit the structural properties of real-world legacy systems. On edge cases, experts appreciated the
resulting programs but noted that some legacy patterns were still missing
despite matching metrics, a limitation of pattern coverage rather than of the
methodology itself. As future work, we plan to scale the pipeline further and
extend our reference corpus to other business domains.
%futur work
% scale the pipeline to generate more programs
% extending the reference corpus to other business domains

\textbf{Acknowledgements.} This work is supported by the Inria Défi LLM4Code (\textcolor{blue}{\url{https://project.inria.fr/llm4code/}}).